\documentstyle[twocolumn,aps]{revtex}

\newcommand{\mb}{\begin{equation}}
\newcommand{\me}{\end{equation}}
\begin{document}
\draft
\title{Classical double-layer atoms: artificial molecules}
\author{B. Partoens\cite{bart}, V. A. Schweigert\cite{vitaly} 
and F. M. Peeters\cite{francois}}
\address{Departement Natuurkunde,
Universiteit Antwerpen (UIA), B-2610 Antwerpen, Belgium}
\date{\today}
\maketitle
\begin{abstract}
The groundstate configuration and the eigenmodes of two parallel
two-dimensional classical atoms are obtained as function of the
inter-atomic distance ($d$). The classical particles are confined by 
identical
harmonic wells and repel each other through a Coulomb potential. As
function of $d$ we find several structural transitions which are of first
or second order. For first (second) order transitions the first (second)
derivative of the energy with respect to $d$ is discontinuous, the radial
position of the particles changes discontinuously (continuously) and the
frequency of the eigenmodes exhibit a jump (one mode becomes soft,
i.e. its frequency becomes zero). 
\end{abstract}


In recent years, there has been an increased interest in the study of
finite two-dimensional (2D) systems of charged
particles~\cite{recent}. These quantum dots are atomic like structures
which have 
interesting optical properties and may be of interest for single electron 
devices.  
Most of the previous works have treated the quantum mechanical problem of a 
small number of electrons. If the number of electrons is increased
beyond 6-7 the Coulomb correlation of the electrons has to be
treated in an approximate way. 
Therefore, in order to fully understand the physics of Coulomb
correlations, we have made a throughout study of the {\it classical
system}~\cite{bedanov} in
which 
the particles are taken point-like but where no 
approximation was made 
on the Coulomb correlation. This approach is valid in high 
magnetic  fields where the kinetic energy of the electrons is quenched. 
In the ground state the charged particles are located on rings and a table of Mendeljev
was constructed in Ref.~\cite{bedanov}.
These configurations were recovered
in a quantum calculation in the limit of large magnetic
fields~\cite{maksym}.
A study of the spectral properties of these
classical
systems such as the energy spectrum, the eigenmodes, and the density of
states was made in Ref.~\cite{schweigert}.
Such classical confined systems have been observed in electron dimples on
liquid helium~\cite{leiderer}, in drops of colloidal
suspensions~\cite{hug} and in confined dusty particles~\cite{chiang}.
\par
Here we extend our previous work to the case of classical
{\it artificial molecules}~\cite{ugajin} which consist of two classical 2D atoms
which are laterally separated by a distance $d$. Intuitively, we 
expect interesting behavior as function of $d$ which governs the inter 
atomic interaction. This can be seen as follows, for 
$d=0$ we have just one 2D atom (e.g. for $2N=10$, this is a two ring
structure), while for
$d\rightarrow\infty$ the system consists of two independent 2D atoms with
each half of
the total number of particles (e.g. two atoms each having
5 particles on one ring). This implies that as function of $d$ 
structural transitions (e.g. configurational changes) have to take place.
We found that these artificial molecules show a surprisingly
complex behavior as function of $d$. In this letter, we investigate the
groundstate
energy 
and its derivatives, the groundstate configurations and the normal modes,
i.e. the eigenfrequencies, of the system.
\par

We limit ourselves to the system consisting of an even number, $2N$, of charged
particles
which are evenly distributed over two layers separated a distance $d$.
In both layers, the confinement potential is centered around the $z$-axis 
and this parabolic confinement, which is taken the same for both layers, 
keeps the system
together in the $xy$-plane. We
focus our attention on systems described by the Hamiltonian (where $m$ is
the mass of the particles, $\omega_0$ the radial confinement frequency, $e$
the particle charge, and $\epsilon$ the dielectric constant of the medium
the particles are moving in) 
\begin{equation}
H=H_{I}+H_{II}+H_{I,II},
\end{equation}
with
\begin{mathletters}
\begin{equation}
H_S=\sum_{i\in S}\frac{1}{2}m\omega_o^2r_i^2+\frac{e^2}{\epsilon}
\sum_{i<j \in S}\frac{1}{|\vec{r}_i-\vec{r}_j|},
\end{equation}
the confinement and interaction energy of the artificial atom in layer
$S=I,II$, and
\begin{equation}
H_{I,II}=\frac{e^2}{\epsilon}\sum_{i\in I}\sum_{j\in II}
\frac{1}{|\vec{r}_i-\vec{r}_j|},
\end{equation}
the interaction energy between the atoms in the two layers.
\end{mathletters}
For convenience, we will refer to our charged particles as electrons,
keeping in mind that they can also be ions with charge $e$ and mass $m$.
The Hamiltonian can be written in a dimensionless form if we express the
coordinates, energy, and frequency in the following units
$r'=(e^2/\epsilon)^{1/3}\alpha^{-1/3},
E'=(e^2/\epsilon)^{2/3}\alpha^{1/3},
\omega'=\omega_0/\sqrt{2},$ respectively,
with $\alpha=m\omega_0^2/2$.
The numerical values for the parameters
$r'$ and $E'$ for some typical experimental systems were given in
Ref.~\cite{bedanov}.
In dimensionless units the Hamiltonian becomes
\begin{eqnarray}
H&=&\sum_{i\in I}r_i^2+\sum_{i<j \in I}
\frac{1}{|\vec{r}_i-\vec{r}_j|}+\sum_{i\in II}r_i^2 \nonumber\\
&&+\sum_{i<j \in
II}\frac{1}{|\vec{r}_i-\vec{r}_j|}+\sum_{i\in
I}\sum_{j\in II}\frac{1}{|\vec{r}_i-\vec{r}_j|},
\end{eqnarray}
and consequently the groundstate energy is only a function of the 
number of electrons, $2N$, and the
distance $d$ between the layers. 
\par
The numerical method used in the present study to obtain the groundstate
configuration is based on
the Monte Carlo technique supplemented with
the Newton Method in order to increase the accuracy of the energy of the
groundstate configuration. The latter technique is outlined and compared
with the
Monte Carlo technique in Ref.~\cite{schweigert} and also
yields the eigenfrequencies and the eigenmodes of the groundstate
configuration.

Let us consider first the case of a molecule consisting of six electrons
distributed over two atoms each with three electrons. For $d=0$ this
is a 2D atom with $6$ electrons, of which we know 
the groundstate configuration~\cite{bedanov}, namely
(1,5): $5$ electrons on a ring and one electron in the center
of the ring. In the opposite limit, $d\rightarrow\infty$, we have two
independent 2D
atoms, each consisting of three electrons for which the groundstate
configuration 
consists of one
ring containing three electrons~\cite{bedanov}. Thus, as function of $d$, 
we expect a structural transition.
\par
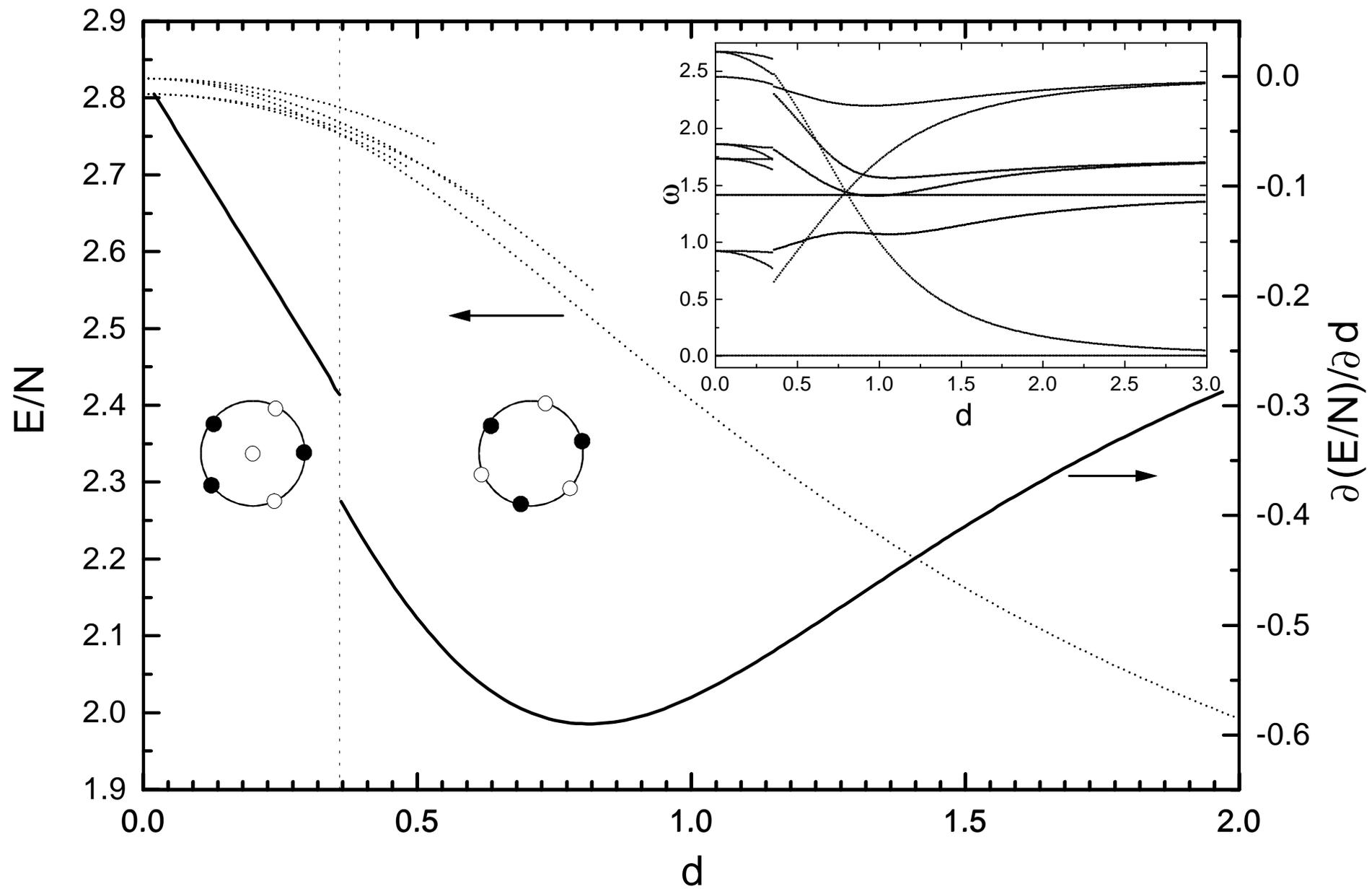
\begin{figure}
\caption{The energy of the groundstate (plotted as dots) 
and its first derivative with
respect to the lateral distance $d$ between the two atoms constituting the
artificial molecule for $2N=6$. The inset shows the frequency of the normal
modes of the groundstate. The configurations are schematically
represented in the figure where solid and open dots correspond to 
electrons in different layers.}
\label{fig1}
\end{figure}

Fig.~\ref{fig1} shows the energy per electron of the groundstate and its
first derivative with respect to $d$. At $d=0.35955$ the first derivative is
discontinuous and a structural transition takes place. For
$d<0.35955$ the groundstate configuration is $(1,2)/(0,3)$: 
the configuration $(1,2)$ in one layer (indicated by the open
dots in the insets of Fig.~\ref{fig1}) and
$(0,3)$ in the other layer (indicated by the solid dots). As viewed from above
we have $(1,5)$ which is 
the configuration of one atom consisting of six electrons.
This implies that the inter-layer correlations are sufficiently strong
to impose the one atom configuration to the $2N=6$ electrons in the 
molecular structure.
For $d>0.35955$ the configuration is twice $(0,3)$, which is the
configuration of two independent atoms each consisting of three electrons.
\par
The frequencies of the normal modes of the groundstate configuration are
shown in the inset of Fig.~\ref{fig1}. Notice that at the first order transition point
the frequencies exhibit a jump.
For $d>0.35955$ more modes are degenerate in energy and in the limit of
$d\rightarrow\infty$ all eigenfrequencies
are at least twofold degenerate. The latter is a consequence of
the fact that the electrons in one layer can vibrate
in phase and out-of-phase with respect to the electrons in the other
layer.  With decreasing $d$ the inter-layer interaction destroys this
degeneracy.
\par
In Fig.~\ref{fig1} also the
energies of
the metastable states are shown. For $d=0$ there is just one metastable
state with
configuration $(0,6)$. If $d$ is slightly different from zero, there are
four metastable states plus the groundstate. One obtains
these
configurations by
moving one electron from one layer to the other without changing its
lateral position, and this for each of the two configurations [(1,5) and
(0,6)].
From 
Fig.~\ref{fig1} one sees clearly how a configuration, metastable for
$d<0.35955$, becomes the stable one for $d>0.35955$. 
\par

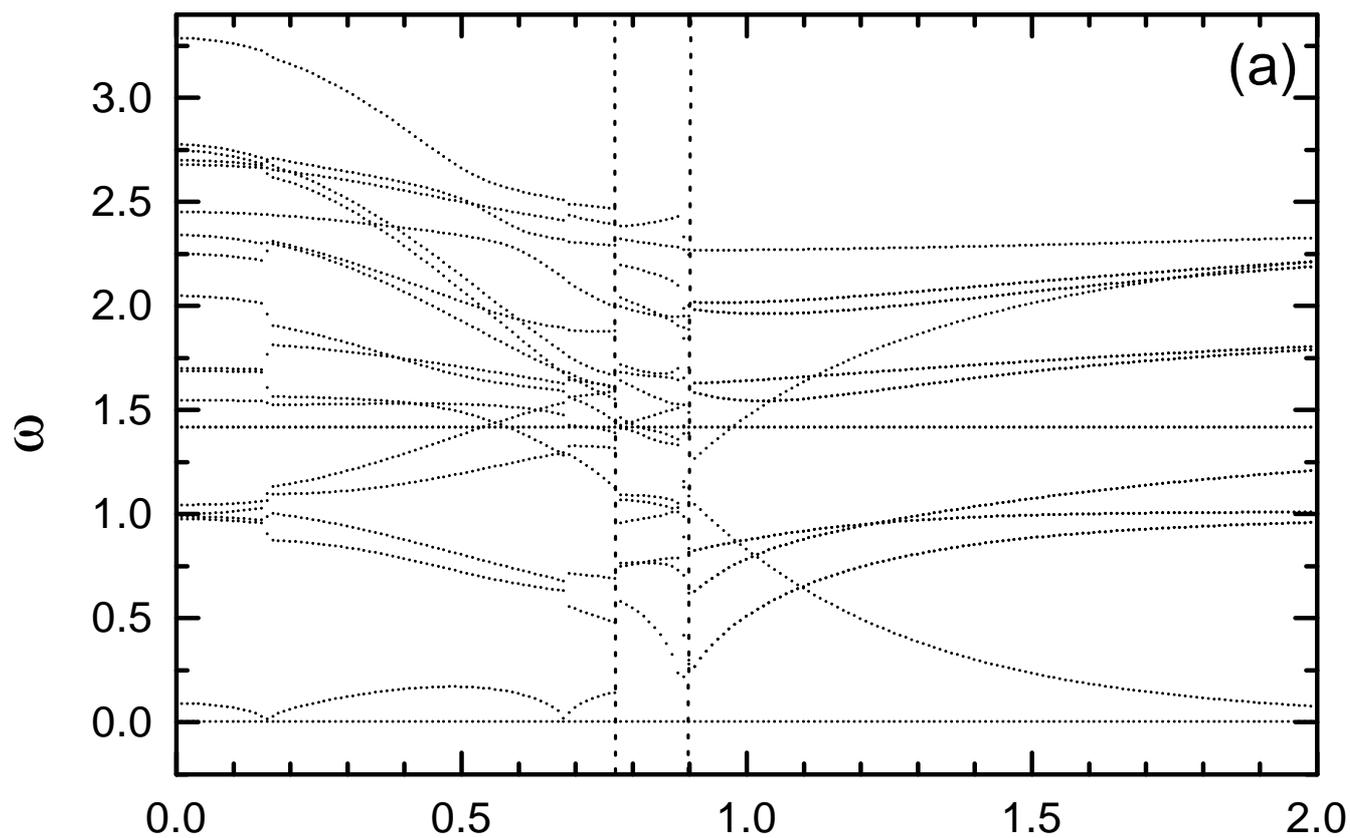
\begin{figure}
\caption{(a) The eigenfrequencies of the groundstate for the artificial
molecule with two times five electrons as a function of the distance $d$ 
between the two atoms. 
(b) The distance of the different electrons from the center of the
confinement potential. The
three groundstate configurations are also shown in the inset of the figure.}
\label{fig3}
\end{figure}

The artificial molecule consisting of two times five electrons is far more
complex and exhibits several transitions some of which are qualitatively
different. For $d=0$ this is a 2D atom with
$10$ electrons and the groundstate configuration is $(2,8)$~\cite{bedanov}. 
For $d\rightarrow\infty$ we have two independent 2D atoms, both with
configuration $(0,5)$~\cite{bedanov}.
Fig.~\ref{fig3}(a) shows the
eigenfrequencies as a function of $d$ and Fig.~\ref{fig3}(b) the
distance of the electrons from the center of the confinement potential. 
From a first glance we have four transition regions,
namely around $d=0.16, d=0.68, d=0.78$ and $d=0.90$.
The spatial configurations of the electrons are depicted in the inset of
Fig.~\ref{fig3}(b).
\par
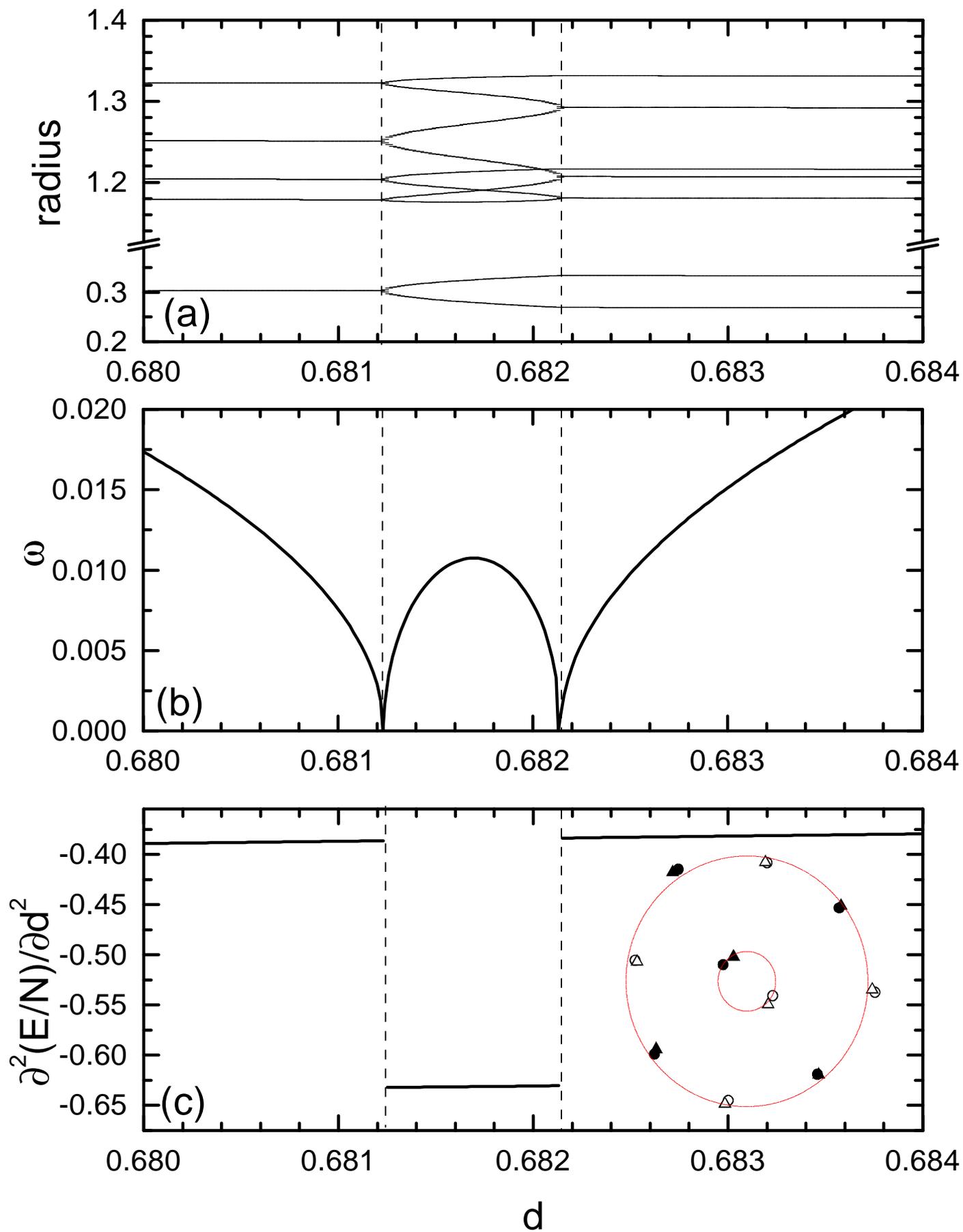
\begin{figure}
\caption{(a) The distance of the electrons from the center of the
confinement
potential for the artificial molecule with two times five electrons as a 
function of $d$. (b) The lowest non zero
eigenfrequency 
of the groundstate. (c) The second derivative
of the groundstate energy
with respect to $d$. The inset shows the configuration at $d=0.681$
(circles) and at $d=0.683$ (triangles). Open and solid symbols refer to
electrons belonging to different atoms.}
\label{fig4}
\end{figure}

\textbf{Region 1} (around $d=0.16$) and \textbf{region 2} (around
$d=0.68$). 
The transitions in both regions are qualitatively similar and therefore
we limit our discussion to $d\approx 0.68$. At these
transition points there is
{\it no abrupt} change of the configuration, but within a 
small $d$-region 
the
radii of the electrons change appreciably, but 
continuously (Fig.~\ref{fig4}(a)).
Fig.~\ref{fig4}(b) shows that this region is delimited by two
eigenfrequencies which become zero and consequently these transitions are
induced by the softening of a mode. In Fig.~\ref{fig4}(c) the
second
derivative of the energy with respect to $d$ is given. There are two
discontinuities in the second derivative, namely at $d=0.68120$
and $d=0.68215$ while the first derivative is continuous. 
In the inset of Fig.~\ref{fig4}(c) the spatial configuration is shown at
$d=0.681$ (circles) and at $d=0.683$ (triangles). The open and
closed symbols refer to electrons belonging to different layers. Notice
that no qualitative changes of the configuration occurs at the second
order transitions. 
For region 1 we found similar second order transitions at $d=0.15775$ and
$d=0.16175$.
\par
\textbf{Region 3} (around $d=0.78$). The transition occurs at
$d=0.77605$ where the position of the electrons changes discontinuously.
There is a jump in the eigenfrequencies and in the first
derivative of the energy. Clearly this is a first order transition which
is similar to those found for the above molecule with two times three
electrons.
\par
\textbf{Region 4} (around $d=0.90$). In this region there are two
transitions.
The first at $d=0.88585$, and the second at $d=0.90588$. Only in the latter
the number of electrons on the different rings changes. It is a first order transition
and the eigenfrequencies exhibit a jump. On the other hand the transition
at
$d=0.88585$ does not correspond to a change in the number of rings,
however there is a jump in
the position of the electrons. It is a first order transition with a
discontinuity in the frequency spectrum
in which the smallest non zero eigenfrequency 
decreases for
$d\rightarrow 0.88585$ but stays different from zero.
\par
For the artificial molecule in which each atom contains six
electrons we find both
first and second order transitions. With increasing $d$ there are 
three first order
transitions, namely at $d=0.2965, d=0.5125$ and $d=0.5345$, followed by
one second order transition induced by the softening of a mode at
$d=0.9259$. The latter is different from the previous molecule with two
times five electrons
where there are 
narrow $d$-regions delimited by two second order transitions.
For $d<0.2965$ the configuration is $(2,4)/(1,5)$, at
the
first order transitions the position of the particles changes
discontinously while at the second order transition the position of
the particles changes continuously and after it the outer
electrons form a perfect circle with the inner electrons sitting in the center,
resulting in the configuration $(1,5)/(1,5)$.
\par
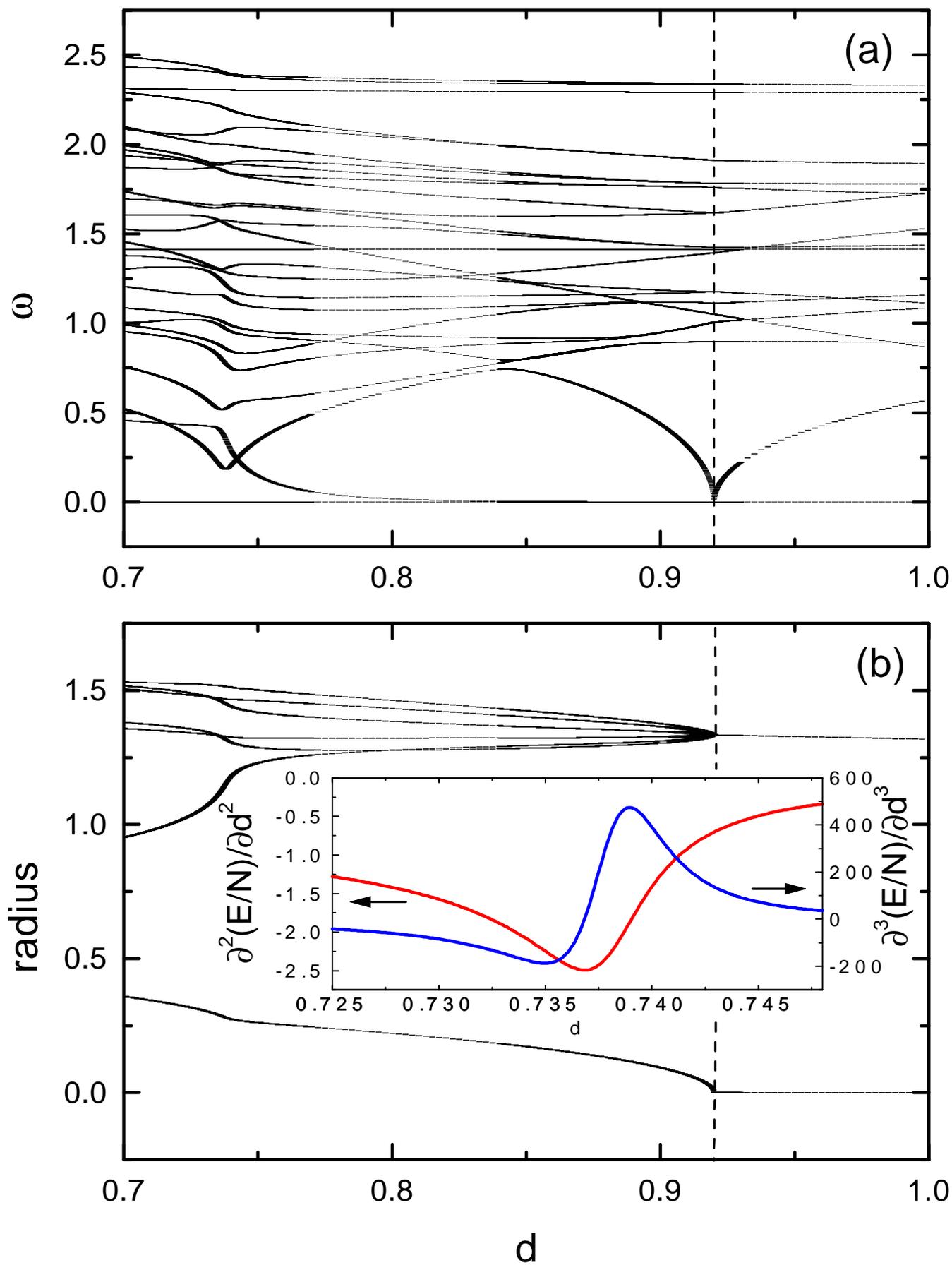
\begin{figure}
\caption{(a) The eigenfrequencies of the groundstate for the artificial
molecule with two times seven electrons as a
function of $d$. (b) The distance of the electrons from the center of
the confinement potential. In the inset the second and third derivative 
of the energy are shown around $d\approx 0.74$.}
\label{fig5}
\end{figure}

The groundstate~\cite{bedanov} of the artificial atom with seven electrons is
$(1,6)$, and with eight electrons is
$(1,7)$.
Therefore, we expect that the above new type of transition also takes
place for
the
artificial
molecule with two times seven and two times eight electrons. For the
molecule
in which both atoms contain seven electrons we find two narrow $d$-regions
delimited by
two second order transitions (the first at $d=0.360375$ and $d=0.360495$,
the second at $d=0.504625$ and $d=0.504865$), a first order transition at
$d=0.5425$, and indeed the second order transition at $d=0.9259$.
Fig.~\ref{fig5} shows the frequency spectrum and the
position of the particles near the last transition. At the second order
transition at $d=0.9259$ the system transforms from an imperfect ring
structure with the middle electron outside the center of the system to 
a perfect ring with one electron in the center of each layer (see
Fig.~\ref{fig5}(b)).
\par
For the latter molecule
another phenomenon occurs. Around $d\approx 0.74$ the smallest non zero eigenfrequency
decreases substantially but does not become zero while the other
eigenfrequencies and the position of the particles
changes fast but continuously with $d$. It is clear
that
this is not a first, second
or third order transition as is made clear in the inset of
Fig.~\ref{fig5}(b) where we plotted the second and third derivative of
the energy with respect to $d$. Therefore we refer to this
as a continuous transition with a lambda-like shape of the derivative 
$\partial^3 E/\partial d^3$. 
\par

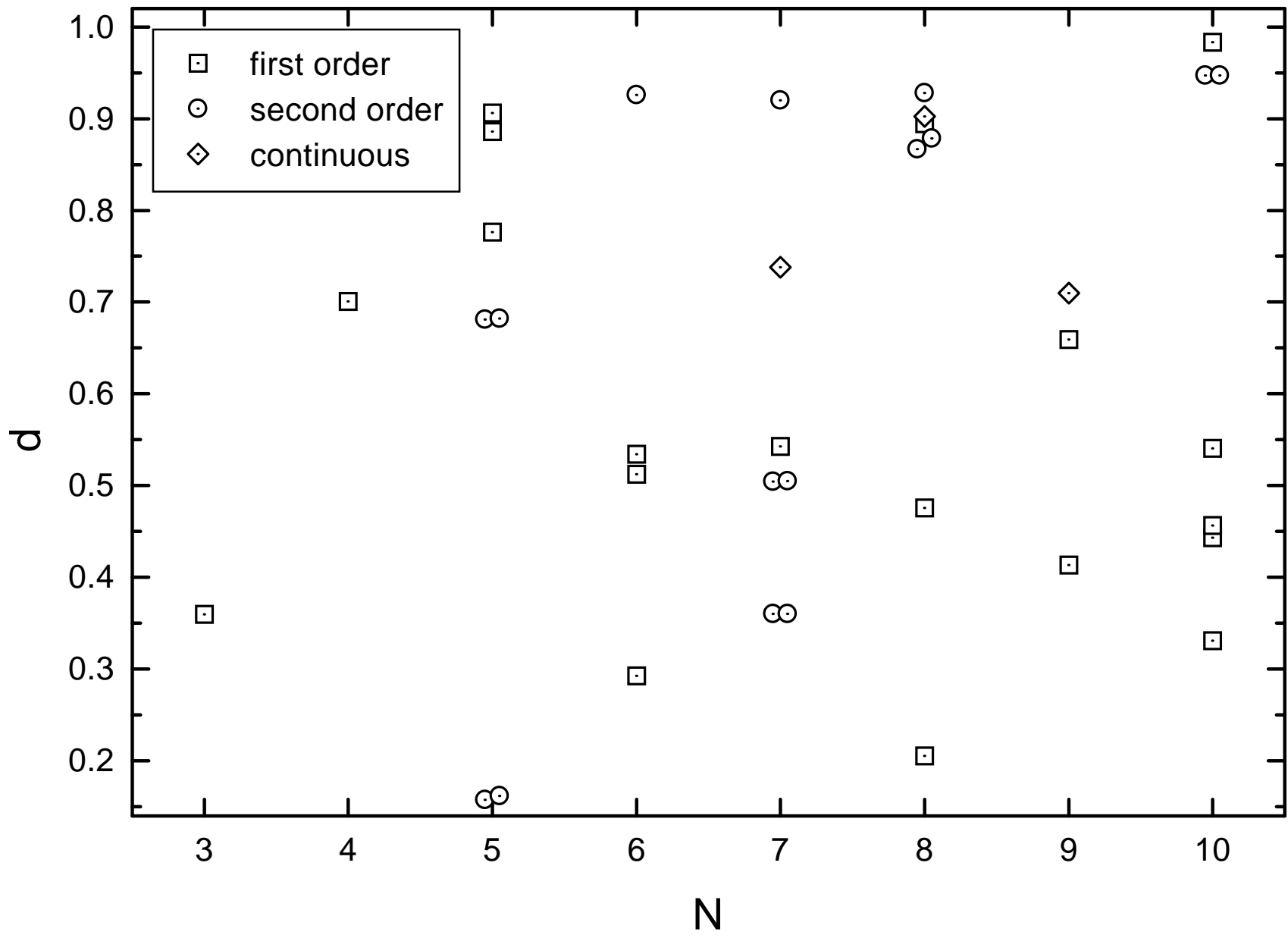
\begin{figure}
\caption{Summary of the different transitions for molecules up to
$2N=20$ electrons. `Continuous' refers to a transition at which the
position and the eigenfrequencies change rapidly, but
continuously, with $d$ and the derivatives of the energy with respect 
to $d$ are
continuous.} 
\label{fig6} 
\end{figure}

In conclusion, we have presented the results of a numerical calculation of the groundstate
configuration, its energy
and the spectrum of the normal modes of classical double-layer atoms.
There is a vertical Coulomb coupling between the electrons
constituting the molecule. These
artificial molecules undergo structural transitions as function of the
distance between the layers. We found first and second order
transitions. For first order transitions the position of the particles
changes discontinuously and the eigenfrequencies exhibit a jump. For second
order transitions the position of the particles changes continuously 
and these
transitions are induced by the softening of an eigenmode. 
The different transitions are summarized in Fig.~\ref{fig6} in a $(d,N)$
diagram for molecules containing up to $2N=20$ electrons.
Although we assumed that: 1) both atoms of the molecule contain the same
number of electrons, and 2) both atoms have the same confinement
potential, a rich variety of structural transitions are found. It is
expected that if we relax these assumptions new configurations are
possible. 
\par
The present system constitutes a simple non trivial model system
exhibiting a rich variety of structural transitions. This finite system
correctly describes the local properties of structural phase transitions.
Furthermore, we found additionally: 1) transitions which are delimited by
recurring second order transitions, and 2) a lambda type of transition without a
divergence in the second derivative of the energy.
It should be noted that in the limit $N\rightarrow\infty$ we
obtain the system of two parallel classical two-dimensional electron gases,
each of them form a
Wigner crystal which was studied in Ref.~\onlinecite{goldoni}.
\par
\par
Acknowledgments: B.P. is an Aspirant and F.M.P. a Research Director of the
Flemish Science Foundation (FWO-Vlaanderen). This work was supported
by a `Krediet aan Navorsers' from the FWO-Vlaanderen, the Human Capital and
Mobility network Programme No. ERBCHRXT 930374, 
INTAS-93-1495, and the Russian Foundation for Basic Research 
96-02-19134.

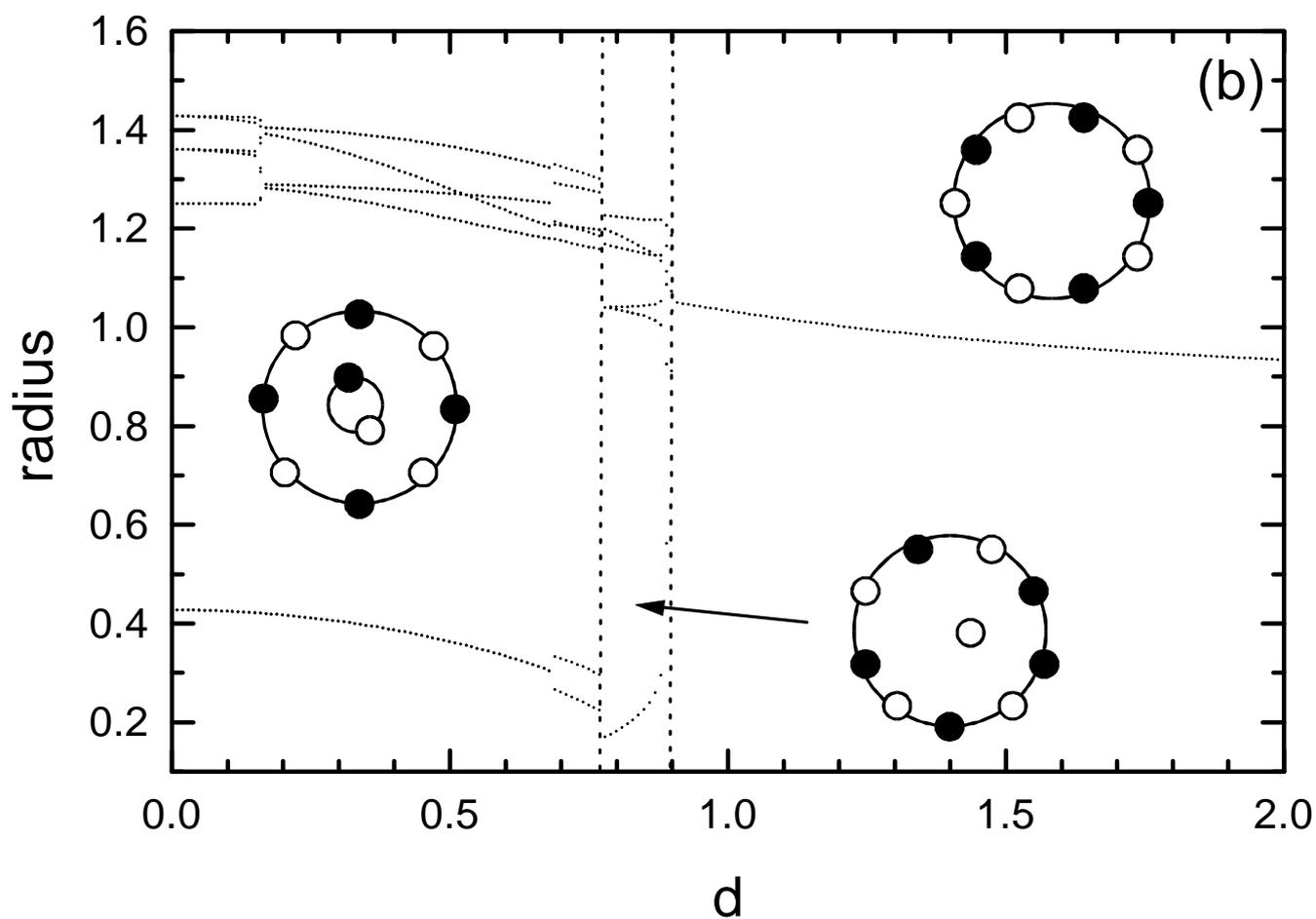
\end{document}